\documentclass{article}

\usepackage{microtype}
\usepackage{graphicx}
\usepackage{subcaption}
\usepackage{booktabs} 

\usepackage{hyperref}

\usepackage[preprint]{icml2026}

\usepackage{amsmath}
\usepackage{amssymb}
\usepackage{mathtools}
\usepackage{amsthm}

\usepackage{xspace}

\newcommand{\pr}{\mathrm{p}}
\newcommand{\f}{\mathrm{f}}

\newcommand{\di}{\mathrm{d}}
\newcommand{\Xc}{\mathcal{X}}
\newcommand{\Oc}{\mathcal{O}}
\newcommand{\Rb}{\mathbb{R}}
\newcommand{\Eb}{\mathbb{E}}
\newcommand{\Pb}{\mathbb{P}}
\newcommand{\Ib}{\mathbb{I}}

\newcommand{\argmin}{\mathrm{argmin}}

\newcommand{\mbar}{\bar{m}}
\newcommand{\ybar}{y}

\newcommand{\alg}{\textsc{lift}\xspace}

\usepackage[capitalize,noabbrev]{cleveref}

\theoremstyle{plain}
\newtheorem{theorem}{Theorem}[section]
\newtheorem{proposition}[theorem]{Proposition}

\theoremstyle{definition}
\newtheorem{definition}[theorem]{Definition}
\newtheorem{assumption}[theorem]{Assumption}
\theoremstyle{remark}

\crefname{theorem}{Theorem}{Theorems}
\crefname{definition}{Definition}{Definitions}
\crefname{assumption}{Assumption}{Assumptions}

\usepackage[textsize=tiny]{todonotes}

\icmltitlerunning{Lifting Biomolecular Data Acquisition}

\begin{document}

\twocolumn[
  \icmltitle{Lifting Biomolecular Data Acquisition}

\icmlsetsymbol{equal}{*}

  \begin{icmlauthorlist}
    \icmlauthor{Eli N. Weinstein}{equal,comp,yyy}
    \icmlauthor{Andrei Slabodkin}{equal,comp}
    \icmlauthor{Mattia G. Gollub}{equal,comp}
    \icmlauthor{Kerry Dobbs}{equal,comp}
    \icmlauthor{Xiao-Bing Cui}{equal,comp}
    \icmlauthor{Fang Zhang}{comp}
    \icmlauthor{Kristina Gurung}{comp}
\icmlauthor{Elizabeth B. Wood}{comp}
\end{icmlauthorlist}

  \icmlaffiliation{yyy}{Department of Chemistry, Technical University of Denmark, Kgs. Lyngby, Denmark}
  \icmlaffiliation{comp}{JURA Bio, Boston, MA, USA}

  \icmlcorrespondingauthor{Eli N. Weinstein}{enawe@dtu.dk}
  \icmlcorrespondingauthor{Elizabeth B. Wood}{ew@jurabio.com}

\icmlkeywords{Machine Learning, ICML}

  \vskip 0.3in
]

\printAffiliationsAndNotice{}  

\begin{abstract}
One strategy to scale up ML-driven science is to increase wet lab experiments' information density.
We present a method based on a neural extension of compressed sensing to function space.
We measure the activity of multiple different molecules simultaneously, rather than individually.
Then, we deconvolute the molecule-activity map during model training.
Co-design of wet lab experiments and learning algorithms provably leads to orders-of-magnitude gains in information density. 
We demonstrate on antibodies and cell therapies.
\end{abstract}

\section{Introduction}

ML holds dramatic potential to automate and accelerate science.
Algorithms may be used to design and learn from complex laboratory experiments, to extract more from information from finite experimental resources \citep{Lindley1956-ig,Chaloner1995-hv,Rainforth2024-cn}.
As integration deepens, it is unlikely that the optimal experimental designs are those that follow conventional human interpretable protocols. 
Rather, machine-readable experiments could carry higher information density.

Here, we study the co-design of algorithms and experiments in the context of high throughput biology.
We consider experiments that synthesize and test the activity of different biomolecules - we focus on proteins, though the methods are more general. 
The aim is to learn a mapping from molecule to activity.
Such maps are critical for molecular design and disease diagnosis \citep{Frazer2021-uq,Watson2023-sp}.
We want experiments and learning algorithms that, together, extract as much information as possible under resource constraints.

The stock blend of experiments and algorithms is designed for supervised learning \cite{Yang2019-mc}.
First, collect $(x,y)$ pairs recording the activity $y$ of different protein amino acid sequences $x$. 
Then, fit a model, such as a neural network, to predict $y$ from $x$. 
This gives an estimate of the sequence-activity map.
The challenge is that the information we gain is bottlenecked by experimental resources: the only way to learn more is to collect more $(x,y)$ pairs.

In this article, we pursue an orthogonal route to increasing information: we increase the number of $x$ per $y$.  
That is, we test more than one $x$ at the same time, obtaining a composite picture of their activity. As a result, we can no longer assign an activity to each sequence by eye.
But we can still learn algorithmically, by deconvoluting the sequence-activity map during training.
Indeed, it turns out we can learn much more than if we tested just one $x$ at a time.

\paragraph{Overview and contributions} Our approach builds on \emph{compressed screens}, which test multiple different molecules $x$ per measurement $y$, then reconstruct activity with a LASSO-style algorithm \citep{Yao2023-hd,Liu2024-zb}.
In this previous work, the goal is to efficiently learn the activity of each molecule within a fixed set. Our goal instead is to train a model that can generalize, and predict the activity of unobserved molecules.

To do so, we \emph{lift} compressed screens from finite to infinite dimensions. Rather than reconstruct a latent vector, we reconstruct a latent function parameterized by a neural network: the molecule-activity map.
To accomplish this, we develop a functional, neural extension of compressed sensing (\Cref{sec:method}).
We realize this extension experimentally and algorithmically (\Cref{sec:experiment-design}).
Theoretically, we prove the experimental process lets us reconstruct the molecule-activity map efficiently, under weak assumptions on the underlying biophysics (\Cref{sec:theory}).
Empirically, we demonstrate the method lets us train more predictive models, compared to if we tested sequences individually (\Cref{sec:empirics}).
We validate with wetlab experiments on generative model-designed antibodies (scFv CAR-T cell therapy constructs), predicting their binding strength against challenging cancer targets (HLA-presented intracellular tumor antigens). 

\section{Lifting Data Acquisition} \label{sec:method}

Our goal is to learn the relationship $\f(x)$ between biomolecules and their activity.

A standard way to achieve this is to first experimentally test different sequences, collecting $(x_1, y_1), \ldots, (x_n, y_n)$, then fit a model $y_i \sim f_\theta(x_i) + \epsilon_i$, where $\epsilon_i$ denotes the experimental noise.
This returns an estimate $f_{\hat \theta} \approx \f$.
The challenge is that experimental costs limit the number of datapoints $n$, and this limits how accurately we can learn $\f$.

Rather than test one sequence at a time, test a mixture of different sequences, $x_{i1}, \ldots, x_{im_i}$.
Now, we observe 
\begin{equation} \label{eqn:mix-observe}
    \ybar_i \sim \sum_{j=1}^{m_i} \f(x_{ij}) + \epsilon_i.
\end{equation} 
We assume here the activity of each sequence contributes additively to the overall outcome \citep{Yao2023-hd,Liu2024-zb}.

How can we learn $\f$ efficiently? 
As usual, we parameterize $\f$ with a neural network $f_\theta$ and fit it to the data. But it is unclear how to choose the mixtures. It is also unclear how to deconvolve each sequence's individual contribution.

\begin{figure*}[t]
\centering
\includegraphics[width=0.8\linewidth]{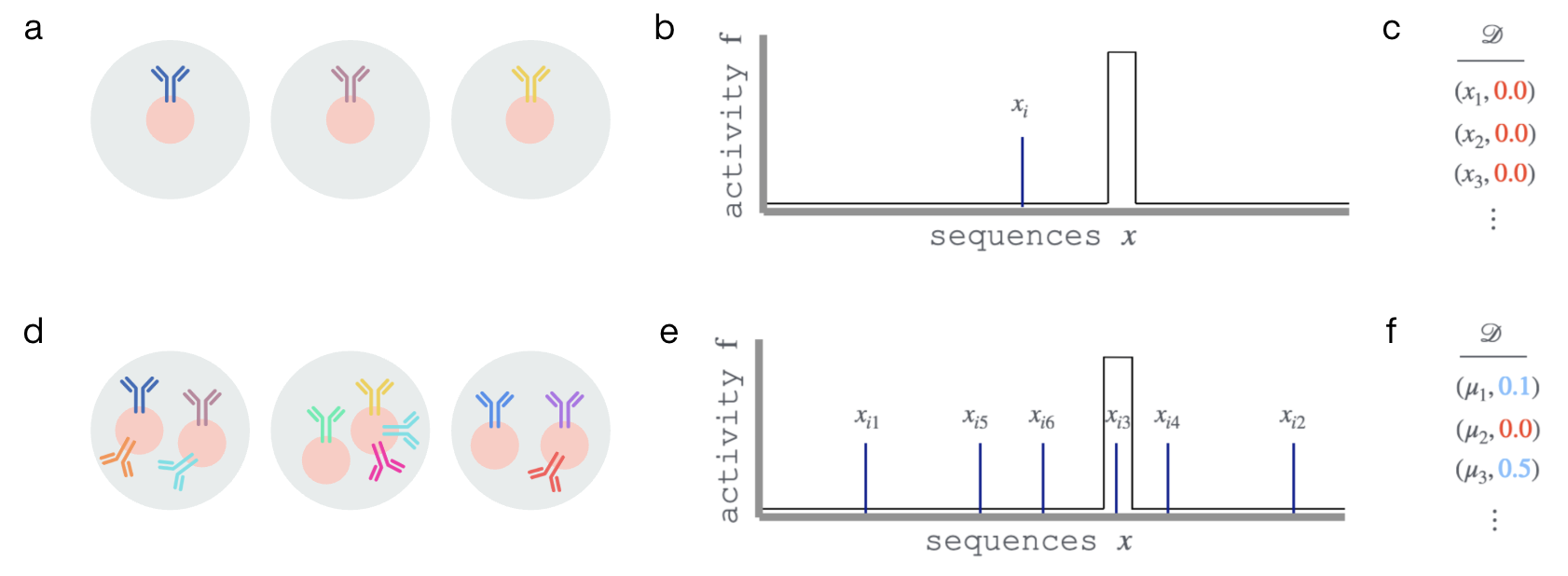}
\caption{\textbf{Lifting data acquisition} Top row: learn from $(x,y)$ pairs. a. Test one sequence per cell. b. When the sequence-activity map $\f$ is sparse (black), most sequences (blue) show no activity. c. So the resulting dataset contains little information. Bottom row: learn from mixtures $\mu_i(x) = \sum_{j=1}^{m_i} \delta_{x_{ij}}(x)$. d. Deliver multiple sequences $x_{ij}$ to each cell. e. With more sequences tested, we are more likely to see signal. f. The resulting dataset provides more information about $\f$, which we deconvolute during training.} \label{fig:overview}
\end{figure*}

\paragraph{Functional, Neural Compressed Sensing}
We have lifted the learning problem from regression on points to regression on mixtures.
Although $\f(x)$ may be arbitrarily complex and nonlinear as a function of $x$, observations are linear in the mixture. To see this, rewrite the mixture as a positive measure, $\mu_i(x) \triangleq \sum_{j=1}^{m_i} \delta_{x_{ij}}(x)$. \Cref{eqn:mix-observe} becomes,
\begin{equation}
	\ybar_i = \f \cdot \mu_i + \epsilon_i,
\end{equation}
where $\f \cdot \mu_i \triangleq \int \f(x) \mu_i(x) \di x$ denotes the $L^2$ inner product on functions. 
So in function space, learning $\f$ is a linear estimation problem.

To take advantage of the linearity, we apply ideas from compressed sensing. 
Compressed sensing designs limited linear measurements to reconstruct high-dimensional signals \citep{Candes2006-qc}. 
It leads us to the following recipe.

\emph{Experimentally, make random incoherent measurements.} If we test one sequence at a time, then $\mu_i(x) = \delta_{x_i}(x)$, and we only learn about the value of $\f$ at a single $x_i$.
This is a maximally \emph{coherent} measurement, meaning $\mu_i(x)$ is concentrated at one point (\Cref{fig:overview}b). It is a very inefficient way to learn a linear model \citep{Candes2011-nc,Russo2018-bn}.
Instead, we should choose $\mu_i$ to be a random mixture of many different sequences, giving information about $\f$ at many locations.
Such $\mu_i(x)$ are \emph{incoherent}, meaning more spread out (\Cref{fig:overview}e). 
Using incoherent measurements reduces the number of datapoints $n$ needed to learn $\f$ \citep{Candes2011-nc}.

\emph{Algorithmically, exploit prior knowledge for regularization.} With limited measurements, the unknown, high-dimensional $\f$ is underdetermined. 
However, we have prior knowledge. 
In biology, sequence-to-activity maps are often sparse: only $x$ within a small region of sequence space show any activity, $\f(x) > 0$ (\Cref{fig:overview}b).
This is particularly true for the hardest and most important design problems, where sequences with a desired activity can be exceedingly rare \citep{Skora2015-qo,Weinstein2025-ei}.
By using an $L^1$ regularizer during training, we can penalize non-sparse $\f$ and resolve the indeterminacy.
The training objective becomes,
\begin{equation}
	\underset{\theta}{\argmin} \sum_{i=1}^n (f_\theta \cdot \mu_i - \ybar_i)^2 + \lambda\|f_\theta\|_{1, \pr} \label{eqn:fLASSO}
\end{equation}
where $\lambda$ is the regularization strength and $\|f\|_{1, \pr} = \int_\Xc |f(x)| \di\pr$ is the $L^1$ norm in function space, with base measure $\pr(x)$ \citep{Benjamin2019-je}.
\Cref{eqn:fLASSO} is the LASSO but with functions in place of vectors. 
The unknown function is parameterized with a neural network.

In summary, we have lifted the learning problem from nonlinear regression on sequences to high-dimensional linear regression on measures.
From this perspective, we see that testing sequences one at a time is inefficient. 
We instead test many, and exploit $\f$'s underlying sparsity to reconstruct the sequence-activity map.

\section{Related Work}

Our method builds on the \emph{lifting trick} in optimization, which equates a non-convex optimization problem, $\max_x \f(x)$, with its convex dual, $\max_{\mu} \f \cdot \mu$ \citep{Bach2021-jq,Yang2020-sq}.
We lift an estimation problem, and take advantage of the fact that measures over molecules can be physically made and tested.

Standard compressed sensing reconstructs vectors. It corresponds to
the case where $f_\theta(x) = \theta_x$ for $\theta \in \Rb^{|\Xc|}$ and $|\Xc| < \infty$ \citep{Donoho2006-wn,Candes2006-bv}.
For example, $x$ may be the row and column of a pixel, and $\theta_x$ its intensity.
Our approach instead applies to structured, high dimensional $x$, such as molecules. It parametrizes $f_\theta(x)$ with a neural network, that can learn features which generalize to unseen $x$.

Previous methods have combined neural networks with compressed sensing, to reconstruct vectors and images. They replace the $\ell^1$ regularizer in the LASSO with a generative model, such as an image generator, and reconstruct a latent representation of the image \citep{Bora2017-lz,Grover2018-zv,Wu2019-tq,Naderi2022-si}. We instead replace the unknown vector with a regression model. 

Previous works have extended compressed sensing from latent vectors to latent functions, to reconstruct images \citep{Adcock2016-qk,Adcock2017-fn}. They work with low-dimensional $x$, e.g. a position in $\Rb^2$ or $\Rb^3$. The unknowns are wavelet coefficients, or the coefficients on another fixed basis, so there is no feature learning.

We build on compressed screens, which apply compressed sensing to biological screening \citep{Yao2023-hd,Liu2024-zb,Cleary2020-zm,Kainkaryam2009-bb}.
These screens test mixtures of biological sequences or small molecules, then reconstruct their activity via the LASSO or a related linear model. 
The success of compressed screens across different molecules and assays helps validate our assumption that outcomes depend additively on the activity of each molecule in a mixture.
However, compressed screening relies on standard vector compressed sensing, describing each molecule's activity with a separate coefficient. 
This prevents generalization and limits scalability: to learn the activity of $x$, you must make multiple measurements of the same $x$.

We employ function-space regularizers and priors developed for neural networks \citep{Fortuin2022-qm,Tran2022-vd}.
Our learning objective in \Cref{eqn:fLASSO} corresponds to maximum \textit{a posteriori} (MAP) estimation under a functional neural network prior \citep{Benjamin2019-je}.
As these previous works point out, it is difficult to design and interpret priors on the parameters of a neural network.
Instead, we use scientific domain knowledge to place a prior in function space.
Unlike many previous neural functional priors, ours is designed to encourage sparsity, rather than smoothness.

\section{Lifting Cell-Based Screens} \label{sec:experiment-design}

We describe an experiment-algorithm pair for high throughput cell-based screens.
In \Cref{sec:theory} we prove it provides sufficiently incoherent experimental measurements to dramatically accelerate learning.

\subsection{Experimental Design}

Our goal is to learn the sequence-activity map $\f$, such that we can make accurate activity predictions for $x \sim \pr(x)$. For example, we may want to accurately predict binding for any human antibody. 
We consider the following laboratory protocol:

\textit{Synthesize sequences.} We synthesize samples from a defined distribution $\pr(x)$ at large scale, using variational synthesis \citep{Weinstein2024-xk}. The distribution $\pr(x)$ can be complex, and specified by a generative protein model. The result is a pool of quadrillions of DNA molecules, each of which is a separate sample $x \sim \pr(x)$.

\textit{Deliver into cells.} Sequences are delivered into cells virally. 
Biophysical models of viral infection describe the number of sequences $m_i$ that enter an individual cell as Poisson, following rare event statistics \citep{Ellis1939-iu}.
The mean of the Poisson is the \emph{multiplicity of infection (MOI)}, $\mbar$. 
We set $\mbar > 1$ to add multiple sequences to each cell, each an independent sample from $\pr(x)$.

\textit{Measure activity.} We use droplet-based single cell sequencing to recover from each cell $i$ both (a) sequences $\{x_{i1}, …, x_{im_i}\}$, and (b) an overall measurement of activity $\ybar_i$. 
For example, $\ybar_i$ can be the number of DNA-barcoded targets that are bound to the cell surface.

The result of this experimental protocol is datapoints $(\{x_{11}, …, x_{1m_1}\}, \ybar_1), \ldots, (\{x_{n1}, …, x_{nm_n}\}, \ybar_n)$, rather than datapoints $(x_1, y_1), \ldots, (x_n, y_n)$.

Another complementary strategy is to \emph{overload} droplets during single cell sequencing, adding multiple cells to each \citep{Datlinger2021-ly,Yao2023-hd,Wu2024-ox}. Since activity is measured at the droplet level, this also provides many $x_{ij}$ per $y_i$.
Droplet overloading can be used separately or together with increased viral MOI. 

\subsection{Model and Training}

From these experiments, we learn a model of $\f$.
We posit the data generating process:
\begin{align}
\theta \sim \pi_\lambda(\theta) = \frac{1}{Z_\lambda} \exp\left(-\lambda \|f_\theta\|_{1, \pr}\right) \label{eqn:dgp_prior}\\
m_i \sim \mathrm{Poisson}(\bar m) \quad \quad \quad x_{ij} \sim \pr(x) \label{eqn:dgp_measurement}\\
\ybar_i \sim \mathrm{Noise}(\sum_{j=1}^{m_i} f_\theta(x_{ij}), \sigma) \label{eqn:dgp_observe}
\end{align}
\Cref{eqn:dgp_prior} is a functional prior over the neural network $f_\theta$. $Z_\lambda$  is its normalizing constant.
\Cref{eqn:dgp_measurement} describes how the measurements $\mu_i$ are generated: $\pr(x)$ is from variational synthesis and the Poisson is from viral delivery at MOI $\bar m$.
So $\mu_i$ is a sample from an inhomogenous Poisson process over sequence space, with rate $\bar m \pr(x)$.
\Cref{eqn:dgp_observe} describes the noisy activity observation. With Gaussian noise, the model's negative log likelihood is directly proportional to the functional neural LASSO in \Cref{eqn:fLASSO}.
For single cell data we use negative binomial noise, parameterized by its mean and dispersion $\sigma$ \cite{Wang2025-rr}. 

We learn $\theta$ by MAP estimation, optimizing the log likelihood of the data. 
To scale to large datasets, we draw minibatches of cells, and optimize $\theta$ with Adam \citep{Kingma2015-ej}.
The term $\|f_\theta\|_{1, \pr}$ involves an intractable integral. We approximate it with Monte Carlo \citep{Benjamin2019-je},
\begin{equation}
	\|f_\theta\|_{1, \pr} = \int |f_\theta(x)| \pr(x) dx \approx \frac{1}{K} \sum_{k=1}^K f_\theta(x^{k})
\end{equation}
where $x^k \sim \pr(x)$ are drawn computationally from the variational synthesis model. 

We refer to the combined laboratory and training procedure as \emph{\alg} since it provides a practical procedure to lift the learning problem from points to measures.

\section{Theory} \label{sec:theory} 

In this section we study \alg theoretically.
We reduce \alg to standard compressed sensing in a simplified setting, such that we can analytically compute the measurements' incoherence.
We find that when $\f$ is sparse, collecting $n$ measurements with \alg can provide as accurate a reconstruction of $\f$ as collecting $\bar m\, n$ conventional $(x,y)$ datapoints.
That is, the effective dataset size scales linearly with MOI.

\subsection{Model Reduction} To tractably analyze \alg, we ignore neural feature learning. We instead consider a simplified model of $\f$ \citep{Lee2020-kg}.
\begin{assumption}[Decision tree approximation]  \label[assumption]{asm:reduced}
\textit{Assume $f_\theta$ takes the form of a decision tree, with $\theta=w \in \Rb^D$ and}
\begin{equation} \label{eqn:reduced}
	f_{w}(x) = \sum_{d=1}^D w_d \mathbb{I}(x \in V_d).
\end{equation}
\textit{Here the $V_d$ form a finite partition of the input space: $\cup_{d=1}^C \cup V_d = \Xc$ and $V_d \cap V_{d'} = \emptyset$ for $d \neq d'$, for $C < \infty$.}
\end{assumption}
For discrete $x$ such as molecules, any $f_\theta$ can be decomposed as \Cref{eqn:reduced}. The simplification is that the features $V_d$ are fixed, rather than learned, $D$ is finite, and the weights $w_d$ are unconstrained.
Intuitively, we can interpret $D$ as the \textit{effective} size of sequence space, or the number of different types of sequences

Under this simplification, \alg reduces to a vector linear model.
The proof, in \Cref{apx:reduction-proof}, relies on the observation that $\mu_i$ is a Poisson process.
\begin{proposition}[Reduction to vector compressed sensing] \label[proposition]{thm:reduce}
	Assume full support, $\pr(x) > 0$ for all $x \in \Xc$. 
Under \Cref{asm:reduced}, the generative process for $y$ (\Cref{eqn:dgp_prior,eqn:dgp_measurement,eqn:dgp_observe}) is equivalent to \begin{align}
		w_d &\sim \mathrm{Laplace}(0, (\lambda \beta_d)^{-1})\\
		a_{id} &\sim \mathrm{Poisson}(\bar m \beta_d)\\
		\ybar_i &\sim \mathrm{Noise}(w \cdot a_{i}, \sigma)
	\end{align}
	where $\beta_d \triangleq \pr(x \in V_d)$. With Gaussian noise, the MAP estimator of $\f_w$ coincides with the LASSO minimizer,
	\begin{equation} \label{eqn:reparam-map}
		\underset{\tilde w \in \Rb^d}{\argmin} \frac{1}{2}\sum_{i=1}^n (\ybar_i - \tilde w \cdot \tilde a_{i})^2 + \tilde \lambda \| \tilde w \|_1
	\end{equation}
	where $\tilde w_d \triangleq \beta_d w_d$, $\tilde a_{id} \triangleq a_{id}/\beta_d$,  $\tilde \lambda \triangleq \lambda \sigma^2 $,  and $\|\tilde w \|_1 = \sum_{d} |\tilde w_d|$ is the $\ell_1$ norm.
\end{proposition}

\subsection{Efficiency Gain} Our goal is now to understand the quality of our experimental design. Does random viral delivery of DNA from a variational synthesis library result in efficient reconstruction of $\f$? What MOI should we use?
To address this, we apply the theory of compressed sensing.
\citet{Candes2011-nc} show that the quality of an experimental design depends critically on the measurements' \emph{coherence}.
\begin{definition}[Coherence [Candes \& Plan, 2011]] \label[definition]{def:coherence}
\textit{Let $a \sim q(a)$ denote a random vector in $\Rb^D$, with zero mean, $\Eb[a] = 0$ and identity covariance, $\Eb[a a^\top] = I_D$. The coherence $\nu_q$ is the smallest number such that
\begin{equation}
    \max_{d}  |a_d|^2 \le \nu_q
\end{equation}
holds deterministically, or stochastically in the sense that
\begin{align}
    &\Eb[D^{-1} \|a\|_2^2 \mathbb{I}(\max_d |a_d|^2 > \nu_q)] \le \frac{1}{20} D^{-3/2}\\& \mathbb{P}(\max_d |a_d|^2 > \nu_q) \le D^{-2},
\end{align}
where the expectation and probability are with respect to $q$.}
\end{definition}
The coherence determines how much data we need to accurately estimate the true model.
\begin{proposition}[Summary of Theorem 1.1-1.2, Candes \& Plan 2011] \label[proposition]{thm:candes-plan}
	Assume a well-specified linear model $y_i = \tilde w \cdot a_i + \epsilon_i$ where $a_i \overset{iid}{\sim} q(a)$ and $\epsilon_i \sim \mathrm{Normal}(0, \sigma)$. 
    Assume $s$ entries of $\tilde w \in \Rb^D$ are non-zero.
    Then if we collect 
    \begin{equation}
        n \ge \nu_q s \log D
    \end{equation}
    datapoints, we can reconstruct $\tilde w$ accurately with high probability using the LASSO. In particular, we obtain an estimate $\hat w$ with $\|\hat w -  \tilde w\|_2^2 \le \mathrm{polylog}(D) \frac{s}{n} \sigma^2 $.
\end{proposition}
We can interpret $n/\nu_q$ as the \emph{effective} dataset size. With less coherence $\nu_q$, we need fewer datapoints $n$ to achieve high accuracy. 
\citet{Candes2011-nc} show the result is quite tight, i.e. we cannot substantially reduce $n$ and still obtain an accurate reconstruction. 
They also show the result is robust if $\tilde w$ is only approximately $s$ sparse, or if the noise is non-Gaussian.

We now examine the coherence of our experimental design.
For simplicity, we assume there is an even probability of a sequence falling in each region $V_d$.
\begin{proposition} \label[proposition]{thm:poisson-cohere}
    Assume $\pr(x \in V_d) = 1/D$ and \Cref{asm:reduced}. If we test sequences $x \sim \pr(x)$ individually, obtaining data $(x_1, y_1), \ldots$, the coherence is $\nu_q = D$. If we use Poisson measurements (\Cref{eqn:dgp_measurement}) then, $\nu_q \le$
\begin{equation} \label{eqn:poisson-cohere}
    16\left(1 + 2 \sqrt{\frac{D}{\bar m}} + \frac{D}{\bar m}\right)\log^2 \delta
\end{equation}
where $\delta = \max(\left(20\sqrt{2}\,D^2\,\sqrt{1 - \frac{2}{D} + \frac{1}{\bar m}}\right)^4, 4 D^6)$
\end{proposition}
Proof in \Cref{apx:poisson-cohere}.
To interpret this result, first note the effective size of sequence space $D$ can be very large. When the MOI $\bar m$ is small compared to $D$, $\nu_q$ is of order $\Oc(\frac{D}{\bar m})$.
Then, increasing the MOI by an order of magnitude, e.g. from 1 to 10, increases our effective data size, $n/\nu_q$, by roughly an order of magnitude. 
That is, testing mixtures of 10 sequences is about as informative as collecting 10 times more $(x,y)$ datapoints.

Once $\bar m \approx D$, further increases in $\bar m$ provide little advantage. At this point, $\nu_q$ is of order $\Oc(\log^2 D)$, versus $\mathcal{O}(D)$ for low MOI. So in the high MOI regime, we effectively obtain orders of magnitude more data.

\section{Empirical Results} \label{sec:empirics}

We examine \alg in simulation and in wet lab experiments, comparing to conventional methods for collecting and training on $(x,y)$ pairs.
Our key finding is that the combination of modified experimental design (high MOI) and modified training algorithm (functional neural LASSO) substantially boosts predictive performance.

\subsection{Synthetic data}

We first examined a synthetic data setting, where the true $\f$ is known. We set $\pr(x)$ to be a variational synthesis model of human antibody CDRH3 sequences \citep{Weinstein2024-xk}. Following previous studies of antibody binding, we set $\f(x)$ to increase from zero in the presence of rare amino acid motifs in $x$ (\Cref{apx:synthetic}) \citep{Akbar2021-mk,Pavlovic2021-sn,Weinstein2025-ei}. 
The simulation is designed to closely follow laboratory screening workflows, including sorting steps and noise processes. 

We generated datasets under different MOI $\bar m$. We then compared the \alg model and training algorithm to a standard model and training algorithm that just uses $(x, y)$ pairs. The architecture (a convolutional neural network) and noise model (negative binomial) were kept the same between the two approaches. 
To handle multiple $x$ with conventional training, we split datapoints $(\{x_{i1}, x_{i2}, \ldots\}, y_i)$ into $(x_{i1}, y_i), (x_{i2}, y_i), \ldots$.
We evaluated each model's predictive performance on held out $(x, f(x))$ pairs. To understand the model's ability to predict rare desired properties, we binarize $\f(x)$ at a threshold and report the area under the precision recall curve.

\begin{figure}
\centering
    \includegraphics[width=\linewidth]{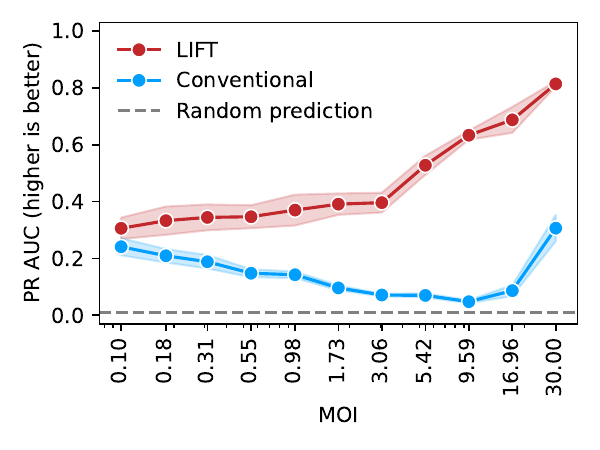}
    \caption{The predictive performance of LIFT training versus conventional training across different experimental designs, on synthetic data. Error bars are s.e. across independent simulations.}
    \label{fig:synthetic_pr_auc}
\end{figure}

The results demonstrate that using LIFT with high MOI produces substantial performance gains (\Cref{fig:synthetic_pr_auc}).
At low MOI, most cells have at most one sequence, so the data is essentially just $(x, y)$ pairs. Here, LIFT training offers minor benefit. But as the MOI increases, LIFT training improves the PR AUC from 0.3 to 0.8, even as conventional training degrades. 
These results demonstrate that co-design of experiments and algorithms is essential: just switching the algorithm, or just increasing MOI, does not lead to performance gains on its own. Instead, it is the combination that provides a major boost in performance.

\subsection{Laboratory experiments: TCRm antibodies} \label{sec:lab-data}

We next deployed LIFT at large scale in the wetlab, with the goal of learning challenging and therapeutically relevant sequence-activity relationships. We focus on binding between human antibodies and peptides displayed in complex with human leukocyte antigens (HLA).
Developing TCR mimicking (TCRm) antibodies that bind specific peptide-HLA (pHLA) complexes could advance oncology by enabling therapeutic targeting of intracellular tumor antigens.
However, discovering such antibodies has been exceptionally difficult \citep{Klebanoff2023-bx}.
This implies the sequence-activity relationship is sparse, making this learning problem a good candidate for LIFT.

We used variational synthesis to create a library of quadrillions of samples from a generative model trained on human post-selection antibody CDRH3s \citep{Weinstein2024-xk}. The library was assembled into scFv CAR cell therapy constructs and delivered into human cells at MOI above one. The library was screened against a panel of fluorescent, DNA-barcoded pHLA targets. 
We sort by fluorescence and use single cell sequencing to recover datapoints $(\{x_{i1}, x_{i2}, \ldots\}, y_i)$ (\Cref{apx:wetlab}) \citep{Weinstein2025-ei}.

\begin{figure}
\centering
    \includegraphics[width=\linewidth]{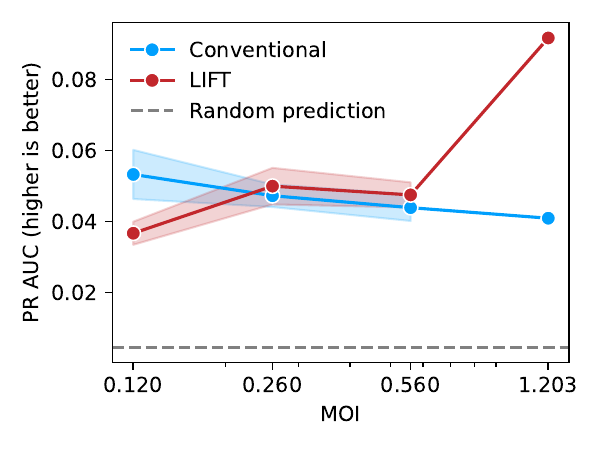}
    \caption{\textbf{Laboratory data: antibody-pHLA binding.} The predictive performance of LIFT training versus conventional training across different experimental designs. Error bars are s.e. across random subsamples of lower MOI datasets.} \label{fig:experimental_pr_auc}
\end{figure}

To evaluate \alg, we designed a controlled internal comparison. 
Our experiment used an MOI just above one, so many cells have just one $x$. This lets us split the data to mimic lower MOI experiments, subsampling cells according to different $m$ distributions. This internal comparison avoids any potential batch variation across experiments.

We trained models to predict binding to MAGE-A4, an oncology target with high tumor specificity that correlates with aggressive disease in a range of solid tumors. 
We evaluated the models on a heldout test set consisting only of cells with $m_i=1$, i.e. $(x, y)$ pairs. We split the data to exclude any $x$ that appeared in the training set, to ensure generalization across sequences.

We find performance on laboratory data closely follow the simulations \Cref{fig:experimental_pr_auc}.
As MOI increases, the performance of conventional training degrades, while LIFT improves.
At the highest MOI, LIFT offers a substantial performance gain over conventional training on low MOI data, increasing the PR AUC by about 50\%.
This implies the model can enrich for hits against MAGE-A4 by a factor of 50\% over the conventional method, while maintaining the same diversity. At some recall values, hits can be enriched by nearly 100\%.
Ablating $L^1$ regularization reduces the PR AUC from $0.092$ to $0.006$.
Overall, we again find that a large performance boost is possible through a modification of both the experimental design and the training algorithm.

\section{Discussion}

We have proposed \alg, a system for packing more information into high throughput biomolecular experiments. 
Theoretical analysis shows \alg can effectively increase the dataset size by orders of magnitude.
When deployed in the wet lab on a challenging therapeutic problem, \alg produces models with 50-100\% improvements in hit rate.

The core idea of \alg is general. We replace supervised learning from $(x, y)$ datapoints with functional estimation from $(\mu, y)$ datapoints. Any experimental setup where multiple inputs contribute additively to the outcome may be lifted in this way. Such situations are abundant in the molecular sciences, where we can physically mix molecules.

\paragraph{Limitations and assumptions}
\alg does not assume the molecule-activity map $\f$ is linear, but it does assume observations are linear: the overall activity depends on the sum of the activity of each molecule. This assumption is violated when one molecule modulates the effect of another, e.g. drug-drug interactions.
\alg also assumes sparsity: molecules with non-zero activity are rare. 
This assumption is violated when there is more gradation in the outcome.
Even when linearity and sparsity are satisfied, \alg may be limited. Using too high an MOI can lead to toxicity and cell death, capping improvements.

\paragraph{Future directions} Going forward, \alg can be extended, its assumptions relaxed, and its robustness improved.  
There are many methods to robustify linear models or incorporate interactions \citep{Buja2019-ur,Filzmoser2021-vu,Prasad2020-wb}. They might be extended from the conventional setting of vector linear models, to \alg's setting of functional neural linear models.
For high dimensional $y$, such as vectors of gene expression, linearity may be more plausible with carefully learned representations \citep{Lotfollahi2019-hn,Wang2023-km}.
Meanwhile, advances in compressed sensing have shown that the sparsity assumption can be replaced with alternative prior assumptions.
For example, one approach is to assume the hidden vector is drawn from a generative model \citep{Bora2017-lz,Grover2018-zv,Wu2019-tq,Naderi2022-si}.
Going forward, this approach might also be extended from vectors to neural functions.

Broadly, our work illustrates a path to scaling up biomolecular ML that does not depend just on added cost.
We can instead pack and extract more information from wet lab experiments, by designing for algorithmic interpretation.

\bibliography{references}

\begin{thebibliography}{45}
\providecommand{\natexlab}[1]{#1}
\providecommand{\url}[1]{\texttt{#1}}
\expandafter\ifx\csname urlstyle\endcsname\relax
  \providecommand{\doi}[1]{doi: #1}\else
  \providecommand{\doi}{doi: \begingroup \urlstyle{rm}\Url}\fi

\bibitem[Adcock \& Hansen(2016)Adcock and Hansen]{Adcock2016-qk}
Adcock, B. and Hansen, A.~C.
\newblock Generalized sampling and infinite-dimensional compressed sensing.
\newblock \emph{Found. Comut. Math.}, 16\penalty0 (5):\penalty0 1263--1323,
  October 2016.

\bibitem[Adcock et~al.(2017)Adcock, Hansen, Poon, and Roman]{Adcock2017-fn}
Adcock, B., Hansen, A.~C., Poon, C., and Roman, B.
\newblock Breaking the coherence barrier: A new theory for compressed sensing.
\newblock \emph{For. Math. Sigma}, 5\penalty0 (e4), 2017.

\bibitem[Akbar et~al.(2021)Akbar, Robert, Pavlović, Jeliazkov, Snapkov,
  Slabodkin, Weber, Scheffer, Miho, Haff, Haug, Lund-Johansen, Safonova,
  Sandve, and Greiff]{Akbar2021-mk}
Akbar, R., Robert, P.~A., Pavlović, M., Jeliazkov, J.~R., Snapkov, I.,
  Slabodkin, A., Weber, C.~R., Scheffer, L., Miho, E., Haff, I.~H., Haug, D.
  T.~T., Lund-Johansen, F., Safonova, Y., Sandve, G.~K., and Greiff, V.
\newblock A compact vocabulary of paratope-epitope interactions enables
  predictability of antibody-antigen binding.
\newblock \emph{Cell Rep.}, 34\penalty0 (11):\penalty0 108856, March 2021.

\bibitem[Bach(2021)]{Bach2021-jq}
Bach, F.
\newblock Finding global minima with kernel approximations, 2021.

\bibitem[Benjamin et~al.(2019)Benjamin, Rolnick, and Kording]{Benjamin2019-je}
Benjamin, A.~S., Rolnick, D., and Kording, K.
\newblock Measuring and regularizing networks in function space.
\newblock In \emph{International Conference on Learning Representations
  (ICLR)}, 2019.

\bibitem[Bora et~al.(2017)Bora, Jalal, Price, and Dimakis]{Bora2017-lz}
Bora, A., Jalal, A., Price, E., and Dimakis, A.~G.
\newblock Compressed sensing using generative models.
\newblock In \emph{International Conference on Machine Learning (ICML)}, March
  2017.

\bibitem[Buja et~al.(2019)Buja, Brown, Berk, George, Pitkin, Traskin, Zhang,
  and Zhao]{Buja2019-ur}
Buja, A., Brown, L., Berk, R., George, E., Pitkin, E., Traskin, M., Zhang, K.,
  and Zhao, L.
\newblock Models as approximations {I}: Consequences illustrated with linear
  regression.
\newblock \emph{Statistical Science}, 34\penalty0 (4):\penalty0 523--544, 2019.

\bibitem[Candes(2006)]{Candes2006-qc}
Candes, E.
\newblock Compressive sampling.
\newblock In \emph{Proceedings of the International Congress of
  Mathematicians}, 2006.

\bibitem[Candes \& Plan(2011)Candes and Plan]{Candes2011-nc}
Candes, E.~J. and Plan, Y.
\newblock A probabilistic and {RIPless} theory of compressed sensing.
\newblock \emph{IEEE Trans. Inf. Theory}, 2011.

\bibitem[Candes \& Tao(2006)Candes and Tao]{Candes2006-bv}
Candes, E.~J. and Tao, T.
\newblock Near-optimal signal recovery from random projections: Universal
  encoding strategies?
\newblock \emph{IEEE Trans. Inf. Theory}, 52\penalty0 (12):\penalty0
  5406--5425, December 2006.

\bibitem[Canonne(2019)]{Canonne2019-ru}
Canonne, C.
\newblock A short note on poisson tail bounds, 2019.

\bibitem[Chaloner \& Verdinelli(1995)Chaloner and Verdinelli]{Chaloner1995-hv}
Chaloner, K. and Verdinelli, I.
\newblock Bayesian experimental design: A review.
\newblock \emph{Stat. Sci.}, 10\penalty0 (3):\penalty0 273--304, 1995.

\bibitem[Cleary \& Regev(2020)Cleary and Regev]{Cleary2020-zm}
Cleary, B. and Regev, A.
\newblock The necessity and power of random, under-sampled experiments in
  biology.
\newblock \emph{arXiv [q-bio.QM]}, December 2020.

\bibitem[Datlinger et~al.(2021)Datlinger, Rendeiro, Boenke, Senekowitsch,
  Krausgruber, Barreca, and Bock]{Datlinger2021-ly}
Datlinger, P., Rendeiro, A.~F., Boenke, T., Senekowitsch, M., Krausgruber, T.,
  Barreca, D., and Bock, C.
\newblock Ultra-high-throughput single-cell {RNA} sequencing and perturbation
  screening with combinatorial fluidic indexing.
\newblock \emph{Nat. Methods}, 18\penalty0 (6):\penalty0 635--642, June 2021.

\bibitem[Donoho(2006)]{Donoho2006-wn}
Donoho, D.~L.
\newblock Compressed sensing.
\newblock \emph{IEEE Trans. Inf. Theory}, 52\penalty0 (4):\penalty0 1289--1306,
  April 2006.

\bibitem[Ellis \& Delbrück(1939)Ellis and Delbrück]{Ellis1939-iu}
Ellis, E.~L. and Delbrück, M.
\newblock The growth of bacteriophage.
\newblock \emph{J. Gen. Physiol.}, 22\penalty0 (3):\penalty0 365--384, January
  1939.

\bibitem[Filzmoser \& Nordhausen(2021)Filzmoser and
  Nordhausen]{Filzmoser2021-vu}
Filzmoser, P. and Nordhausen, K.
\newblock Robust linear regression for high‐dimensional data: An overview.
\newblock \emph{Wiley Interdiscip. Rev. Comput. Stat.}, 13\penalty0 (4), July
  2021.

\bibitem[Fortuin(2022)]{Fortuin2022-qm}
Fortuin, V.
\newblock Priors in bayesian deep learning: A review.
\newblock \emph{Int. Stat. Rev.}, 90\penalty0 (3):\penalty0 563--591, December
  2022.

\bibitem[Frazer et~al.(2021)Frazer, Notin, Dias, Gomez, Min, Brock, Gal, and
  Marks]{Frazer2021-uq}
Frazer, J., Notin, P., Dias, M., Gomez, A., Min, J.~K., Brock, K., Gal, Y., and
  Marks, D.~S.
\newblock Disease variant prediction with deep generative models of
  evolutionary data.
\newblock \emph{Nature}, 599\penalty0 (7883):\penalty0 91--95, 2021.

\bibitem[Grover \& Ermon(2018)Grover and Ermon]{Grover2018-zv}
Grover, A. and Ermon, S.
\newblock Uncertainty autoencoders: Learning compressed representations via
  variational information maximization.
\newblock \emph{AISTATS}, 89:\penalty0 2514--2524, December 2018.

\bibitem[Kainkaryam \& Woolf(2009)Kainkaryam and Woolf]{Kainkaryam2009-bb}
Kainkaryam, R.~M. and Woolf, P.~J.
\newblock Pooling in high-throughput drug screening.
\newblock \emph{Curr. Opin. Drug Discov. Devel.}, 12\penalty0 (3):\penalty0
  339--350, May 2009.

\bibitem[Kingma \& Ba(2015)Kingma and Ba]{Kingma2015-ej}
Kingma, D.~P. and Ba, J.
\newblock Adam: A method for stochastic optimization.
\newblock In \emph{ICLR}, 2015.

\bibitem[Klebanoff et~al.(2023)Klebanoff, Chandran, Baker, Quezada, and
  Ribas]{Klebanoff2023-bx}
Klebanoff, C.~A., Chandran, S.~S., Baker, B.~M., Quezada, S.~A., and Ribas, A.
\newblock {T} cell receptor therapeutics: immunological targeting of the
  intracellular cancer proteome.
\newblock \emph{Nat. Rev. Drug Discov.}, pp.\  1--22, October 2023.

\bibitem[Lee \& Jaakkola(2020)Lee and Jaakkola]{Lee2020-kg}
Lee, G.-H. and Jaakkola, T.~S.
\newblock Oblique decision trees from derivatives of {ReLU} networks.
\newblock In \emph{International Conference on Learning Representations
  (ICLR)}, 2020.

\bibitem[Linderman(2025)]{Linderman2025-tj}
Linderman, S.
\newblock {STATS305B}: Applied statistics {II}, 2025.

\bibitem[Lindley(1956)]{Lindley1956-ig}
Lindley, D.~V.
\newblock On a measure of the information provided by an experiment.
\newblock \emph{Ann. Math. Stat.}, 27\penalty0 (4):\penalty0 986--1005,
  December 1956.

\bibitem[Liu et~al.(2024)Liu, Kattan, Mead, Kummerlowe, Cheng, Ingabire, Cheah,
  Soule, Vrcic, McIninch, Triana, Guzman, Dao, Peters, Lowder, Crawford, Amini,
  Blainey, Hahn, Cleary, Bryson, Winter, Raghavan, and Shalek]{Liu2024-zb}
Liu, N., Kattan, W.~E., Mead, B.~E., Kummerlowe, C., Cheng, T., Ingabire, S.,
  Cheah, J.~H., Soule, C.~K., Vrcic, A., McIninch, J.~K., Triana, S., Guzman,
  M., Dao, T.~T., Peters, J.~M., Lowder, K.~E., Crawford, L., Amini, A.~P.,
  Blainey, P.~C., Hahn, W.~C., Cleary, B., Bryson, B., Winter, P.~S., Raghavan,
  S., and Shalek, A.~K.
\newblock Scalable, compressed phenotypic screening using pooled perturbations.
\newblock \emph{Nat. Biotechnol.}, pp.\  1--13, October 2024.

\bibitem[Lotfollahi et~al.(2019)Lotfollahi, Wolf, and Theis]{Lotfollahi2019-hn}
Lotfollahi, M., Wolf, F.~A., and Theis, F.~J.
\newblock {scGen} predicts single-cell perturbation responses.
\newblock \emph{Nat. Methods}, 16\penalty0 (8):\penalty0 715--721, August 2019.

\bibitem[Naderi \& Plan(2022)Naderi and Plan]{Naderi2022-si}
Naderi, A. and Plan, Y.
\newblock Sparsity-free compressed sensing with applications to generative
  priors.
\newblock \emph{IEEE J. Sel. Areas Inf. Theory}, 3\penalty0 (3):\penalty0
  493--501, September 2022.

\bibitem[Pavlović et~al.(2021)Pavlović, Scheffer, Motwani, Kanduri, Kompova,
  Vazov, Waagan, Bernal, Costa, Corrie, Akbar, Al~Hajj, Balaban, Brusko,
  Chernigovskaya, Christley, Cowell, Frank, Grytten, Gundersen, Haff, Hovig,
  Hsieh, Klambauer, Kuijjer, Lund-Andersen, Martini, Minotto, Pensar, Rand,
  Riccardi, Robert, Rocha, Slabodkin, Snapkov, Sollid, Titov, Weber, Widrich,
  Yaari, Greiff, and Sandve]{Pavlovic2021-sn}
Pavlović, M., Scheffer, L., Motwani, K., Kanduri, C., Kompova, R., Vazov, N.,
  Waagan, K., Bernal, F. L.~M., Costa, A.~A., Corrie, B., Akbar, R., Al~Hajj,
  G.~S., Balaban, G., Brusko, T.~M., Chernigovskaya, M., Christley, S., Cowell,
  L.~G., Frank, R., Grytten, I., Gundersen, S., Haff, I.~H., Hovig, E., Hsieh,
  P.-H., Klambauer, G., Kuijjer, M.~L., Lund-Andersen, C., Martini, A.,
  Minotto, T., Pensar, J., Rand, K., Riccardi, E., Robert, P.~A., Rocha, A.,
  Slabodkin, A., Snapkov, I., Sollid, L.~M., Titov, D., Weber, C.~R., Widrich,
  M., Yaari, G., Greiff, V., and Sandve, G.~K.
\newblock The {immuneML} ecosystem for machine learning analysis of adaptive
  immune receptor repertoires.
\newblock \emph{Nature Machine Intelligence}, 3\penalty0 (11):\penalty0
  936--944, November 2021.

\bibitem[Prasad et~al.(2020)Prasad, Suggala, Balakrishnan, and
  Ravikumar]{Prasad2020-wb}
Prasad, A., Suggala, A.~S., Balakrishnan, S., and Ravikumar, P.
\newblock Robust estimation via robust gradient estimation.
\newblock \emph{J. R. Stat. Soc. Series B Stat. Methodol.}, 82\penalty0
  (3):\penalty0 601--627, July 2020.

\bibitem[Rainforth et~al.(2024)Rainforth, Foster, Ivanova, and
  Smith]{Rainforth2024-cn}
Rainforth, T., Foster, A., Ivanova, D.~R., and Smith, F.~B.
\newblock Modern bayesian experimental design.
\newblock \emph{Stat. Sci.}, 2024.

\bibitem[Russo \& Van~Roy(2018)Russo and Van~Roy]{Russo2018-bn}
Russo, D. and Van~Roy, B.
\newblock Learning to optimize via information-directed sampling.
\newblock \emph{Oper. Res.}, 66\penalty0 (1):\penalty0 230--252, February 2018.

\bibitem[Skora et~al.(2015)Skora, Douglass, Hwang, Tam, Blosser, Gabelli, Cao,
  Diaz, Papadopoulos, Kinzler, Vogelstein, and Zhou]{Skora2015-qo}
Skora, A.~D., Douglass, J., Hwang, M.~S., Tam, A.~J., Blosser, R.~L., Gabelli,
  S.~B., Cao, J., Diaz, Jr, L.~A., Papadopoulos, N., Kinzler, K.~W.,
  Vogelstein, B., and Zhou, S.
\newblock Generation of {MANAbodies} specific to {HLA}-restricted epitopes
  encoded by somatically mutated genes.
\newblock \emph{Proc. Natl. Acad. Sci. U. S. A.}, 112\penalty0 (32):\penalty0
  9967--9972, August 2015.

\bibitem[Tran et~al.(2022)Tran, Rossi, Milios, and Filippone]{Tran2022-vd}
Tran, B.-H., Rossi, S., Milios, D., and Filippone, M.
\newblock All you need is a good functional prior for bayesian deep learning.
\newblock \emph{J. Mach. Learn. Res.}, January 2022.

\bibitem[Wang et~al.(2025)Wang, Shu, Cao, and Grima]{Wang2025-rr}
Wang, Y., Shu, Z., Cao, Z., and Grima, R.
\newblock From noise to models to numbers: Evaluating negative binomial models
  and parameter estimations in single-cell {RNA}-seq.
\newblock \emph{bioRxiv}, pp.\  2025.05.05.652189, May 2025.

\bibitem[Wang et~al.(2023)Wang, Gui, Negrea, and Veitch]{Wang2023-km}
Wang, Z., Gui, L., Negrea, J., and Veitch, V.
\newblock Concept algebra for (score-based) text-controlled generative models.
\newblock \emph{arXiv [cs.CL]}, February 2023.

\bibitem[Watson et~al.(2023)Watson, Juergens, Bennett, Trippe, Yim, Eisenach,
  Ahern, Borst, Ragotte, Milles, Wicky, Hanikel, Pellock, Courbet, Sheffler,
  Wang, Venkatesh, Sappington, Torres, Lauko, De~Bortoli, Mathieu, Ovchinnikov,
  Barzilay, Jaakkola, DiMaio, Baek, and Baker]{Watson2023-sp}
Watson, J.~L., Juergens, D., Bennett, N.~R., Trippe, B.~L., Yim, J., Eisenach,
  H.~E., Ahern, W., Borst, A.~J., Ragotte, R.~J., Milles, L.~F., Wicky, B.
  I.~M., Hanikel, N., Pellock, S.~J., Courbet, A., Sheffler, W., Wang, J.,
  Venkatesh, P., Sappington, I., Torres, S.~V., Lauko, A., De~Bortoli, V.,
  Mathieu, E., Ovchinnikov, S., Barzilay, R., Jaakkola, T.~S., DiMaio, F.,
  Baek, M., and Baker, D.
\newblock De novo design of protein structure and function with {RFdiffusion}.
\newblock \emph{Nature}, 620\penalty0 (7976):\penalty0 1089--1100, August 2023.

\bibitem[Weinstein et~al.(2024)Weinstein, Gollub, Slabodkin, Gardner, Dobbs,
  Cui, Amin, Church, and Wood]{Weinstein2024-xk}
Weinstein, E.~N., Gollub, M., Slabodkin, A., Gardner, C., Dobbs, K., Cui, X.,
  Amin, A.~N., Church, G.~M., and Wood, E.~B.
\newblock Manufacturing-aware generative protein models enable {DNA} synthesis
  of samples at petascale.
\newblock 2024.

\bibitem[Weinstein et~al.(2025)Weinstein, Slabodkin, G~Gollub, and
  B~Wood]{Weinstein2025-ei}
Weinstein, E.~N., Slabodkin, A., G~Gollub, M., and B~Wood, E.
\newblock Accelerated learning on large scale screens using generative library
  models.
\newblock \emph{arXiv [stat.ML]}, October 2025.

\bibitem[Wu et~al.(2024)Wu, Bennett, Ye, Sridhar, Eidenschenk, Everett,
  Nazarova, Chen, Kim, Deangelis, Owen, Chen, Lau, Shi, Lund,
  Xavier-Magalhães, Patel, Liang, Modrusan, and Darmanis]{Wu2024-ox}
Wu, B., Bennett, H.~M., Ye, X., Sridhar, A., Eidenschenk, C., Everett, C.,
  Nazarova, E.~V., Chen, H.-H., Kim, I.~K., Deangelis, M., Owen, L.~A., Chen,
  C., Lau, J., Shi, M., Lund, J.~M., Xavier-Magalhães, A., Patel, N., Liang,
  Y., Modrusan, Z., and Darmanis, S.
\newblock Overloading and {unpacKing} ({OAK}) - droplet-based combinatorial
  indexing for ultra-high throughput single-cell multiomic profiling.
\newblock \emph{Nat. Commun.}, 15\penalty0 (1):\penalty0 9146, October 2024.

\bibitem[Wu et~al.(2019)Wu, Rosca, and Lillicrap]{Wu2019-tq}
Wu, Y., Rosca, M., and Lillicrap, T.
\newblock Deep compressed sensing.
\newblock In \emph{International Conference on Machine Learning (ICML)}, May
  2019.

\bibitem[Yang et~al.(2019)Yang, Wu, and Arnold]{Yang2019-mc}
Yang, K.~K., Wu, Z., and Arnold, F.~H.
\newblock Machine-learning-guided directed evolution for protein engineering.
\newblock \emph{Nat. Methods}, 16\penalty0 (8):\penalty0 687--694, August 2019.

\bibitem[Yang et~al.(2020)Yang, Zhang, Chen, and Wang]{Yang2020-sq}
Yang, Z., Zhang, Y., Chen, Y., and Wang, Z.
\newblock Variational transport: A convergent {particle-based} algorithm for
  distributional optimization.
\newblock \emph{arXiv [cs.LG]}, December 2020.

\bibitem[Yao et~al.(2023)Yao, Binan, Bezney, Simonton, Freedman, Frangieh, Dey,
  Geiger-Schuller, Eraslan, Gusev, Regev, and Cleary]{Yao2023-hd}
Yao, D., Binan, L., Bezney, J., Simonton, B., Freedman, J., Frangieh, C.~J.,
  Dey, K., Geiger-Schuller, K., Eraslan, B., Gusev, A., Regev, A., and Cleary,
  B.
\newblock Scalable genetic screening for regulatory circuits using compressed
  perturb-seq.
\newblock \emph{Nat. Biotechnol.}, October 2023.

\end{thebibliography}
\bibliographystyle{icml2026}

\newpage
\appendix
\onecolumn
\section{Proof of \Cref{thm:reduce}}\label{apx:reduction-proof}

We first rewrite the prior (\Cref{eqn:dgp_prior}). We have
\begin{equation}
	\|f_w\|_{1, \pr} = \int |f_w(x)| \pr(x) dx = \int \sum_{d=1}^D |w_d| \mathbb{I}(x \in V_d) p(x) = \sum_{d=1}^D |w_d| \beta_d.
\end{equation}
So the prior becomes
\begin{equation}
	\pi_\lambda(w) = \frac{1}{Z_w} \prod_{d=1}^D \exp(-\lambda \beta_d |w_d|)
\end{equation}
and so $w_d \sim \mathrm{Laplace}(0, (\lambda \beta_d)^{-1})$.

Next we rewrite the measurement process (\Cref{eqn:dgp_measurement}). Recall $\mu_i = \sum_{j=1}^{m_i} \delta_{x_{ij}}(x)$. Then $\mu_i(x)$ is a sample from an inhomogenous Poisson process over sequence space $\Xc$ with intensity $\bar m \pr(x)$ \citep[e.g.][]{Linderman2025-tj}.
Define,
\begin{equation}
	a_{id} \triangleq \int_{x \in V_d} \mu_i(x) dx 
\end{equation}
Since the $V_d$ are disjoint subsets of $\Xc$, we find from the properties of Poisson processes that $a_{id} \sim \mathrm{Poisson}(\bar m \pr(x \in V_d))$ independently.

Finally, we rewrite the mean of the observation process (\Cref{eqn:dgp_observe}) as
\begin{equation}
	\sum_{j=1}^{m_i} f_w (x_{ij}) = f_w \cdot \mu_i = \sum_{d=1}^D w_d \int_{x \in V_d} \mu_i(x) dx = \sum_{d=1}^D w_d a_{id}
\end{equation}

We have derived the first part of \Cref{thm:reduce}. 

With Gaussian noise, the MAP estimate of $w$ is,
\begin{equation}
	\hat{w} \triangleq\underset{w \in \Rb^d}{\argmin} \sum_{i=1}^n \frac{1}{2 \sigma^2} (y - w \cdot a_i)^2 + \sum_{d=1}^D \lambda \beta_d |w_d| = \underset{w \in \Rb^d}{\argmin} \sum_{i=1}^n \frac{1}{2} (y - w \cdot a_i)^2 + \lambda \sigma^2 \sum_{d=1}^D  \beta_d |w_d|
\end{equation}
We can reparameterize this optimization problem to work with $\tilde w \triangleq w \odot \beta$ where $\odot$ denotes elementwise multiplication. This gives \Cref{eqn:reparam-map}. Then, from the solution $\hat{\tilde{w}}$ to \Cref{eqn:reparam-map}, we obtain $\hat w = \hat{\tilde{w}} \odot \beta^{-1}$, where $\beta^{-1}_d = 1/\beta_d$. 
To ensure this is well defined, we need $\beta_d > 0$ for all $d$. This is guaranteed by the assumption that $\pr(x) > 0$.

\section{Proof of \Cref{thm:poisson-cohere}} \label{apx:poisson-cohere}

First assume we collect data about individual sequences $x \sim \pr(x)$ instead of mixtures. Then we have, from the same argument as in \Cref{apx:reduction-proof}, that $a_{i} \sim \mathrm{Categorical}(1/D, \ldots, 1/D)$. \citet{Candes2011-nc} show that this has coherence $\nu_q = D$.

Next we consider the coherence of the Poisson measurements (\Cref{thm:reduce}). To compute the coherence, we must first write the measurements in the standardized form. Define $r = \bar m / D$ and define $q(\tilde a)$ as,
\begin{align}
a_{id} \overset{iid}{\sim} \mathrm{Poisson}(r) \quad \quad 
    \tilde a_{id} \triangleq \frac{a_{id} - r}{\sqrt r}.
\end{align}
We see that $\tilde a$ has mean zero and covariance the identity.
Note that to understand the performance of \alg under \Cref{thm:reduce} it suffices to analyze the coherence of $\tilde a$, since the outcome $y$ is linear in $\tilde a$.

Now we compute the coherence of $\tilde a$. First, via a union bound,
\begin{align}
    \Pb(\max_d |\tilde A|^2 > \nu) \le& D \Pb(|\tilde A| > \sqrt \mu)\\
    & =D \Pb(|A_{id} - r| > \sqrt{r \nu})
\end{align}
Using the subexponential bound on the Poisson from \citet[Theorem 1]{Canonne2019-ru},
\begin{align}
    &\le 2 D \exp(-\frac{r \nu}{2 (r + \sqrt{r \nu})})\\
    &\le 2 D \exp(-\frac{\sqrt{r \nu}}{2 (\sqrt{\frac{r}{\nu}} + 1)})\\
    &\le 2 D \exp(-\frac{\sqrt{r \nu}}{2 (\sqrt{r} + 1)}) \label{eqn:inter-poisson-bound}
\end{align}
where in the last line we use $\nu \ge 1$ \citep{Candes2011-nc}.

We require $\Pb(\max_d |\tilde A|^2 > \nu) \le D^{-2}$, which we can now see holds so long as
\begin{equation} \label{eqn:bound-pt1}
    \nu_q \ge \left(1 + \frac{1}{\sqrt{r}}\right)^2\log^2 \left(4 D^6\right)
\end{equation}

Next we consider the second bound determining the coherence. From Cauchy-Schwarz,
\begin{align}
    &\Eb[D^{-1} \|\tilde A\|_2^2 \Ib(\max_d |\tilde A_d|^2 > \nu)]\\
    &\le \sqrt{\Eb[D^{-2} (\|\tilde A\|_2^2)^2] \Eb[\Ib(\max_d |\tilde A_d|^2 > \nu)]}\\
    &=\sqrt{\frac{1}{D^2}(D \Eb[\tilde A^4] + (D^2 - D) \Eb[\tilde A^2]^2) \Pb(\max_d |\tilde A_d|^2 > \nu)}\\
    &\le \sqrt{2(D - 2 + \frac{1}{r})} \exp(-\frac{\sqrt{r \nu}}{4 (\sqrt{r} + 1)})
\end{align}
where in the last line we plugged in \Cref{eqn:inter-poisson-bound} and the kurtosis of a Poisson distribution, $1/r + 3$.

We require $\Eb[D^{-1} \|\tilde A\|_2^2 \Ib(\max_d |\tilde A_d|^2 > \nu)] \le \frac{1}{20}D^{-3/2}$.
This implies
\begin{equation}
    \nu \ge 16 \left(1 + \frac{1}{\sqrt{r}}\right)^2 (\log 20 + \frac{1}{2} \log 2 + \frac{3}{2} \log D + \frac{1}{2} \log(D - 2 + \frac{1}{r}))^2
\end{equation}
Simplifying, \begin{equation}
    \nu \ge 16 \left(1 + \frac{1}{\sqrt{r}}\right)^2 \log^2(20 \sqrt 2 D^2(1 - \frac{2}{D}  + \frac{1}{r D})^{1/2})
\end{equation}
The result follows, after recalling that $rD = \bar m$.

\section{Synthetic Data} \label{apx:synthetic}

In brief, our synthetic data simulations follow the same setup as in \citet{Weinstein2025-ei}, except that instead of providing each cell just one $x$, we sample $m_i$ values of $x_{ij}$, with $m_i$ drawn from a Poisson (\Cref{eqn:dgp_measurement}). 

As in \citet{Weinstein2025-ei}, $\f(x)$ is a deterministic function (based on the presence of specific amino acid sequences at specific positions) and we add noise to generate $y$. In particular, $y$ follows a Negative Binomial distribution parametrized by deterministic 'sequence strength' $\f(x)$. By default, $\f(x) = 1$, and the presence of any of the following patterns increases $\f(x)$ by a multiplier of 10: (i) \texttt{"P"} or \texttt{"C"} at position 3 (zero-indexed), (ii) \texttt{"N"} or \texttt{"C"} at position 5, and (iii) "\texttt{"PC"} or \texttt{"SS"} starting at position 6. The final observable value of $y$ is sampled from the Negative Binomial distribution with a mean of $\f(x)$ and variance $\f(x) + \frac{\f(x)^2}{\phi}$ (where $\phi$ = 2.28). 

Then, for each MOI value in $0.1,  0.18,  0.31,  0.55,  0.98,  1.73,  3.06,  5.42,  9.59, 16.96, 30.0$, we sample $N$ ($N = 2000$) cells, drawing $m_i$ from a Poisson with $\lambda=$MOI. For each cell, we sample the corresponding number of sequences from $p(x)$ (which is, same as in \citet{Weinstein2025-ei}, a variational synthesis model of human antibody CDRH3 sequences \citep{Weinstein2024-xk}) along with their observable strengths as described above. Again, as in \citet{Weinstein2025-ei}, we assume only hits are sequenced, so we select only the cells with total observable strength $>= 10$ and employ a LeaVS correction (\Cref{apx:leavs}). Note that $N$ is fixed, which yields different numbers of non-zero counts and, hence, hits, for different MOIs. We repeated the experiment $10$ times, each time generating the data independently. We used the CNN-based model architecture described in \citet{Weinstein2025-ei}.
We evaluate precision recall in terms of the model's ability to predict whether $\f(x)$ is greater than 10.

\section{LeaVS} \label{apx:leavs}

We derive a LeaVS correction term following \cite{Weinstein2025-ei}. After sorting, we do not sequence cells with $y =0$, but know the number of such cells. From \Cref{eqn:dgp_measurement} and \Cref{eqn:dgp_observe} we can compute the corresponding likelihood term as
\begin{equation}
    \pr(y=0) = \sum_{m=0}^\infty \left[\int\pr(y=0 \mid f(x))\pr(x) dx\right]^m \mathrm{Poisson}(m \mid \bar m)
\end{equation}
where $\pr(y=0 \mid f(x))$ is the likelihood under the noise model. We can estimate this term by Monte Carlo.

During training, we add the term $n_0 \log \pr(y=0)$ to our log likelihood, where $n_0$ is the number of cells with $y=0$. 

\paragraph{Connection to $L^1$ functional regularization}

Although the motivation is different, the LeaVS log likelihood correction implicitly applies functional $L^1$ regularization. To see this, note that with Poisson noise, $y \sim \mathrm{Poisson}(f_\theta(x))$, and $m=1$, the LeaVS correction becomes
\begin{align}
    \log \int p_\theta(y = 0 \mid x) \pr(x) = \log \int \exp(-|f_\theta(x)|) \pr(x) dx \label{eqn:leavs_correction}
\end{align}
By Jensen's inequality, this upper bounds $-\int |f_\theta(x)|\pr(x) dx$, the $L^1$ regularizer. In practice, we must approximate the integral in \Cref{eqn:leavs_correction} with finite samples, which makes the bound even tighter. Indeed, since $f_\theta(x)$ is typically close to zero in practice, Taylor approximation shows 
\begin{align}
    \log \int \exp(-|f_\theta(x)|) \pr(x) dx &\approx \log(1 - \int |f_\theta(x)|p(x)dx)\\
    &\approx -\int |f_\theta(x)|\pr(x)dx
\end{align}

In short, the LeaVS correction approximates the $L^1$ regularizer.
So to ablate $L^1$ regularization in \Cref{sec:empirics}, we drop the LeaVS correction.

\section{Laboratory Experiments} \label{apx:wetlab}

Laboratory experiments on TCRms follow \citep{Weinstein2024-xk} (forthcoming manuscript), with higher viral titer. 

To make sure there is no shared information between the training and the test set, we split the data such that no cell and no sequence is present in both sets. To achieve this, we first constructed a bipartite graph where each cell and each sequence is represented by a vertex, and a cell-vertex and a sequence-vertex are connected if this sequence is present in this cell, and then we split the connected components of this graph. As a result, there were 8704 cells in the training set and 976 in the test set. 
We evaluate performance on the test set in terms of PR AUC by thresholding, setting cells with more than $y=10$ counts to be hits.

To perform the comparison for varying MOI, we (i) used the full dataset described above and (ii) subsampled the full training set to create lower-MOI sets. We kept the test set constant for all MOI for comparability. The subsamples were created as follows: first, we estimated the MOI of the full dataset as $1.203$. Then, we draw cells based on this MOI, following the assumption that, conditional on the number of sequences that actually occur in the cell, the MOI $\bar m$ does not affect the distribution of $x$ and $y$. 

In detail, since only the hits were sequenced, we first use a truncated Poisson to estimate the MOI $\bar m$. Then we assume that if we had used low MOI, our dataset would consist of $\tilde n$ samples from the above distribution, where $\tilde n = N \mathrm{Poisson}(m > 0 \mid \bar m ) \textsc{hit-rate}$ with $N$ the total number of cells and \textsc{hit-rate} the fraction of all cells that are hits ($y > 0$). To create a dataset with a lower MOI $\bar m'$, we subsampled cells following the truncated Poisson distribution with $\lambda = \bar m'$: the fraction of cells with $k$ sequences was set to $\mathrm{Poisson}(m = k \mid \bar m' , m>0)$, and the total number of hits was $\tilde n' = N \mathrm{Poisson}(m > 0 \mid \bar m' ) \textsc{hit-rate}$. We repeated the above procedure $10$ times for each MOI $\in [0.12, 0.26, 0.56]$.

\end{document}